\documentstyle[aps]{revtex}

\begin{document}
\title{$1/N$ Expansion for Critical Exponents of Magnetic Phase Transitions in $%
CP^{N-1}$ Model at $2<d<4$}
\author{V.Yu.Irkhin, A.A.Katanin and M.I.Katsnelson}
\address{Institute of Metal Physics, 620219 Ekaterinburg, Russia}
\maketitle

\begin{abstract}
Critical exponents in the $CP^{N-1}$ model, which describes localized-moment
ferro- and antiferromagnets ($N=2$ in the Heisenberg model), are calculated
from two-particle Green's functions to first order in $1/N$. For $%
d=2+\varepsilon $ the results agree with earlier renormalization group
calculations. For $d=3$ the leading $1/N$-corrections turn out to be very
large at $N=2.$ For $d=4-\varepsilon $ the $1/N$ -corrections are small at
any $N$ and insufficient to describe correctly the magnetic phase transition.
\end{abstract}


Recently the large-$N$ approach has been extensively applied to investigate
thermodynamic properties of localized-moment magnets \cite{ArovasBook}. Two
ways of constructing the large-$N$ approach are possible: in the $O(N+1)$
(which is a generalization of the nonlinear sigma model to the case $N>2$)
and in the $SU(N)$-model. At $N=2$ these models describe the usual
Heisenberg magnet and coincide because of existence of homomorphism of $%
SU(2) $ on $O(3)$ group:

\begin{equation}
{\bf n}=z^{\dagger }{\bf \sigma }z\,  \label{Tr}
\end{equation}
where $z$ is the two-component complex field with the constraint condition$%
\, $%
\begin{equation}
\,|z|^2=1  \label{con}
\end{equation}
${\bf n}$ is the corresponding element of $O(3)$ group, ${\bf \sigma }$ are
the Pauli matrices. At the same time, for $N>2$ such an equivalence does not
hold, and two models under consideration correspond to two different
pictures of the localized-magnet excitation spectrum. The $N\rightarrow
\infty $ limit in the $O(N)$ model corresponds to the quantum spherical
model \cite{SphModel}, and the $1/N$ corrections are small \cite{Hikami2}.
However the approach \cite{Arovas} which appeared to be rather successful
for a wide range of problems concerning low-dimensional magnets (see e.g.%
\cite{Yoshioka,Takahashi,Our1,Our2}) is based on the $SU(N)$ rather than $%
O(N)$ model. The $N\rightarrow \infty $ limit in the $SU(N)$-model
corresponds to the mean-field approximation (the modified spin-wave theory) 
\cite{Arovas}.This approximation was successfully applied to two-dimensional
Heisenberg model \cite{Arovas,Yoshioka} and lead to the result for
correlation length, which was consistent with the one-loop scaling analysis 
\cite{Halperin}. Corrections of the first order in $1/N$ \cite{Starykh}
modify only preexponential factor in this case. But when applied to the
Heisenberg model with $d>2$, where a finite phase transition temperature
exists, the mean-field approximation leads to some difficulties \cite{Our1}.
In particular, it yields values of critical exponents $\beta =1,\eta =4-d,$
which are far (for not small $d-2$) from the values obtained from the $1/N$
expansion in the $O(N)$ model \cite{Hikami2}.

In the present paper we investigate $1/N$ corrections to the above-mentioned
critical exponents of the mean-field approximation in the $SU(N)$ model
describing quantum magnets. In the continuum limit this model reduces to the 
$CP^{N-1}$ model \cite{Read}. The latter model enables one to use standard
field theory methods for treating static and dynamical properties in the
vicinity of the magnetic phase transition.

To calculate the whole set of critical exponents we have to consider the
non-uniform longitudinal susceptibility which may be represented in terms of
a two-particle Green's function. The one- and two-particle Green's functions
can be obtained from the generating functional for the $CP^{N-1}$ model \cite
{CP} 
\begin{eqnarray}
Z[h,h^{\dagger }] &=&\int D[z_m^{\dagger },z_m]\delta (|z|^2-1)\exp
(-S[z_m^{\dagger },z_m])  \nonumber \\
S[z_m^{\dagger },z_m] &=&2\rho _s\int\limits_0^{1/T}d\tau \int d^dr\left[
|\partial _\mu z|^2-|z^{\dagger }\partial _\mu z|^2+h^{\dagger
}z+hz^{\dagger }\right]  \label{Z0}
\end{eqnarray}
where $\rho _s$ is the spin stiffness, the magnon velocity is put to be
unity, $z_m$ and $h_m$ are $N$-component complex fields: $m=1...N$.
Representing the delta function in (\ref{Z0}) as a path integral over the
slave field $\lambda $ and decoupling quartic terms in this representation
with the use of the Hubbard-Stratonovich transformation we find the action
in the standard form (see, e.g.,\cite{Starykh})

\begin{equation}
Z[h^{\dagger },h]=\int D[z_m^{\dagger },z_m]DA_\mu D\lambda \exp
(-S[z_m^{\dagger },z_m,A_\mu ,\lambda ])  \label{Z}
\end{equation}
\begin{eqnarray}
S[z_m^{\dagger },z_m,A_\mu ,\lambda ] &=&2\rho _s\int\limits_0^{1/T}d\tau
\int d^dr\left[ |(\partial _\mu -iA_\mu )z|^2\right.  \nonumber \\
&&\ \ \ \left. +i\lambda (|z|^2-1)+h^{\dagger }z+hz^{\dagger }\right]
\label{ZZ}
\end{eqnarray}
To describe the magnetic long-range order we have to take into account the
spontaneous symmetry breaking. We pick up the average part that is supposed
to exist for the field $z_{i1}$: $z_{im}=\overline{z}_m+\widetilde{z}%
_{im},\,\,\overline{z}_m=$ $\langle z_{i1}\rangle \delta _{m,1}.$ We
consider the renormalized classical region, which is sufficient to calculate
critical exponents. Then only the zeroth component in the Matsubara sums
should be retained. Note that the below results for critical exponents are
valid also in the ferromagnetic case since only dynamic part of the action (%
\ref{ZZ}) is different for ferro- and antiferromagnet.

Following to \cite{Starykh} we integrate out the field $\widetilde{z}$ from (%
\ref{Z}), 
\begin{equation}
Z[h^{\dagger },h]=\int DA_\mu D\lambda \exp (NS_{eff})  \label{Z1}
\end{equation}
The effective action takes the form 
\begin{eqnarray}
S_{eff} &=&\frac 2g\int \frac{d^dk}{(2\pi )^d}\int \frac{d^dq}{(2\pi )^d}%
\left( \overline{z}^{\dagger }R_{0k}+h_k^{\dagger }\right) \left( k^2\delta
_{kq}+R_{kq}\right) ^{-1}\left( \overline{z}R_{q0}+h_q\right)  \nonumber \\
&&\ \ \ +\frac 2g\left[ i\lambda _{k=0}-|\,\overline{z}\,|^2R_{00}-(%
\overline{z}^{\dagger }h_{k=0}+\overline{z}h_{k=0}^{\dagger })\right] -\ln
\det_{k_1,k_2}\left[ k_1^2\delta _{k_1k_2}+R_{k_1k_2}\right]  \label{ZZ1}
\end{eqnarray}
with $g=N/\rho _s$ ,

\begin{equation}
R_{k_1k_2}=(k_1+k_2)_\mu A_{k_1-k_2,\mu }+i\lambda
_{k_1-k_2}+\sum_qA_{q-k_1,\mu }A_{k_2-q,\mu }
\end{equation}
The constraint condition (\ref{con}) takes the form 
\begin{equation}
1=\sum_m<z_{im}^{\dagger }z_{im}>=\sum_{qm}\left. \frac{\partial ^2Z}{%
\partial \widetilde{h}_{qm}^{\dagger }\partial \widetilde{h}_{qm}}\right|
_{h=0}  \label{Constr}
\end{equation}
where $\widetilde{h}_{qm}=2\rho _sh_{qm}$. According to the structure of the
transformation (\ref{Tr}), the non-uniform longitudinal susceptibility for $%
N=2$ is expressed as a quartic form in the fields $z$. For an arbitrary $N$
we adopt the following definition: 
\begin{equation}
\chi ^{zz}(q)=\frac 1{2N}\sum_{pkm}\left. \frac{\partial ^4Z}{\partial 
\widetilde{h}_{p-q,m}^{\dagger }\partial \widetilde{h}_{k+q,m}^{\dagger
}\partial \widetilde{h}_{pm}\partial \widetilde{h}_{km}}\right| _{h=0}
\end{equation}
Expanding (\ref{Z1}) in powers of $1/N$ yields a series for $Z$. Below the
magnetic ordering temperature $T_M$ the Bose condensation takes place and
the constraint (\ref{Constr}) in the mean-field approximation $(N\rightarrow
\infty )\,$\cite{Arovas,Yoshioka} takes the form 
\[
n_B=1-\frac{gT}2\int \frac{d^dk}{(2\pi )^d}\frac 1{k^2} 
\]
where the number of condensed bosons $n_B=|\,\overline{z}\,|^2$ plays the
role of the long-range order parameter in the mean-field approximation. The
susceptibility reads 
\begin{equation}
\chi ^{zz}(q)=\frac 1{2N}n_B^2\delta _{q0}+\frac 1{8\rho _s^2}\int \frac{d^dk%
}{(2\pi )^d}\frac 1{k^2(k+q)^2}
\end{equation}
In the next order in $1/N$ we have to introduce \cite{Starykh} inverse
propagator for the field $\lambda $%
\begin{eqnarray}
\Pi _\lambda (q) &=&T\int \frac{d^dk}{(2\pi )^d}\frac 1{k^2(k+q)^2}+\frac{%
4n_B}{gq^2}  \nonumber \\
\ &=&\frac{TK_dA_d}{q^{4-d}}+\frac{4n_B}{gq^2}  \label{PL}
\end{eqnarray}
and spatial part of inverse propagator for the field $A$%
\begin{eqnarray}
\Pi _{ij}(q) &=&2\,T\delta _{ij}\int \frac{d^dk}{(2\pi )^d}\frac 1{k^2}%
-T\int \frac{d^dk}{(2\pi )^d}\frac{(2k+q)_i(2k+q)_j}{k^2(k+q)^2}  \nonumber
\\
&&\ +\frac{4n_B}g(\delta _{ij}-\frac{q_iq_j}{q^2}) 
\begin{array}{c}
=
\end{array}
\Pi _A(q)(\delta _{ij}-\frac{q_iq_j}{q^2})
\end{eqnarray}
\begin{equation}
\Pi _A(q)=\frac{TA_dK_d}{d-1}q^{d-2}+\frac{4n_B}g  \label{PA}
\end{equation}
Here 
\begin{eqnarray}
A_d &=&\frac{\Gamma (d/2)\Gamma (2-d/2)\Gamma ^2(d/2-1)}{2\Gamma (d-2)} 
\nonumber \\
K_d^{-1} &=&2^{d-1}\pi ^{d/2}\Gamma (d/2)
\end{eqnarray}
Note that, unlike the two-dimensional case \cite{Starykh}, the polarization
operator $\Pi _A$ has no ``dangerous'' dependence $q^2$ at small $q$ and
does not lead to infrared divergences.

Diagrams describing $1/N$-corrections to the one-particle Green functions
and non-uniform susceptibility are represented in Fig 1 and Fig.2
respectively. Corresponding analytical form of the constraint equation (\ref
{Constr}) is 
\begin{eqnarray}
1 &=&\frac{gT}2\int \frac{d^dk}{(2\pi )^d}\frac 1{k^2}+\frac{gT}2\int \frac{%
d^dk}{(2\pi )^d}\frac 1{k^4}\left[ \Sigma _A(k)-\Sigma _\lambda (k)\right] 
\nonumber \\
&&\ \ +n_B\left\{ 1-\frac{gT}{2N}\int \frac{d^dk}{(2\pi )^d}\frac 1{\Pi
_\lambda (k)}+g\left[ \frac{\Sigma _A(k)-\Sigma _\lambda (k)}{k^2}\right]
_{k\rightarrow 0}\right\}  \label{Constr1/N}
\end{eqnarray}
Here 
\begin{eqnarray}
\Sigma _\lambda (k) &=&\frac TN\int \frac{d^dq}{(2\pi )^d}\frac 1{\Pi
_\lambda (q)}\left[ \frac 1{(k+q)^2}-\frac 1{k^2}\right]  \nonumber \\
\ &=&\frac{4-d}{d(d-2)}\frac{k^2}{NA_d}L(k,n_B) \\
\Sigma _A(k) &=&\frac{4T}N\int \frac{d^dq}{(2\pi )^d}\frac 1{(k+q)^2}\frac 1{%
\Pi _A(q)}\left[ k^2-\frac{(k_\mu q_\mu )^2}{q^2}\right]  \nonumber \\
\ &=&4\frac{(d-1)^2}{d(d-2)}\frac{k^2}{NA_d}L(k,n_B)
\end{eqnarray}
are the self-energy corrections (see Fig.1) calculated to logarithmic
accuracy and 
\[
L(k,n_B)=\left\{ 
\begin{array}{cc}
\ln (\Lambda ^{d-2}/n_B) & k^{d-2}\ll n_B \\ 
\ln (\Lambda ^{d-2}/k^{d-2}) & k^{d-2}\gg n_B
\end{array}
\right. 
\]
$\Lambda $ being a cutoff parameter. Delta-like contribution to the
susceptibility is obtained by picking out the terms with two condensate
lines in the diagrams. They read 
\begin{eqnarray}
\delta \chi ^{zz}(q) &=&\frac 1{2N}n_B^2\left\{ 1-\frac 2N\int \frac{d^dk}{%
(2\pi )^d}\frac 1{k^4}\frac 1{\Pi _\lambda (k)}\right.  \nonumber \\
&&\ \ \left. +4\left[ \frac{\Sigma _A(k)-\Sigma _\lambda (k)}{k^2}\right]
_{k\rightarrow 0}\right\} \delta _{q0}  \label{Hi1/N}
\end{eqnarray}
It is natural to define the (staggered) magnetization as a square root of
the coefficient at the delta-function in (\ref{Hi1/N}). Note that to first
order in $1/N$ the magnetization does not already coincide with density of
condensed bosons. Evaluating integrals in (\ref{Constr1/N}) and (\ref{Hi1/N}%
)\ yields corrections to the critical exponent $\beta $ due to presence of
singular terms of order of $\ln n_B$. The result is

\begin{equation}
\beta =1-2\frac{d^2-d+2}{NA_d}  \label{Beta}
\end{equation}
Corrections to the staggered susceptibility are determined by diagrams in
Fig. 2. Picking out logarithmically divergent in $q$ terms at $T=T_M$ $%
(n_B=0)$ and $q\rightarrow 0$ leads to the result

\begin{equation}
\eta =(d-2)\left( 1-\frac 8{NA_d}\right)  \label{Eta}
\end{equation}
Using the scaling hypothesis we can calculate other critical exponents. The
results to first order in $1/N\,$are summarized in the Table. \ For $%
d=2+\varepsilon $ we write down two first non-vanishing terms of the
expansion in $\varepsilon $. For comparison, critical exponents to first
order in $1/N$ in the $O(N)$ model \cite{Hikami2,MaBook} and results of the
renormalization group (RG) analysis at $d=2+\varepsilon $ in the nonlinear
sigma model \cite{Brezin} are presented too.

.

Table1. Critical exponents in different models

$
\begin{tabular}{||cc|cc|c|c||}
\hline\hline
& \multicolumn{3}{|c}{$d=2+\varepsilon $} & \multicolumn{2}{|c||}{$d=3$} \\ 
\cline{5-6}\cline{1-4}
\multicolumn{1}{||c|}{} & \multicolumn{1}{c|}{$O(N+1)$} & 
\multicolumn{1}{|c|}{$CP^{N-1}$} & \multicolumn{1}{c|}{RG} & 
\multicolumn{1}{|c|}{$O(N)$} & $CP^{N-1}$ \\ \cline{5-6}\cline{1-4}
\multicolumn{1}{||c|}{$\alpha $} & \multicolumn{1}{c|}{$-2/\varepsilon +1+2/N%
$} & \multicolumn{1}{|c|}{$-2/\varepsilon +1+4/N$} & \multicolumn{1}{c|}{$%
-2/\varepsilon +3$} & \multicolumn{1}{|c|}{$-1+32/\pi ^2N$} & $-1+144/\pi ^2N%
$ \\ \cline{5-6}\cline{1-4}
\multicolumn{1}{||c|}{$\beta $} & \multicolumn{1}{c|}{$1-2\varepsilon /N$} & 
\multicolumn{1}{|c|}{$1-4\varepsilon /N$} & \multicolumn{1}{c|}{$%
1-2\varepsilon $} & $1/2-4/\pi ^2N$ & $1-64/\pi ^2N$ \\ 
\cline{5-6}\cline{1-4}
\multicolumn{1}{||c|}{$\eta $} & \multicolumn{1}{c|}{$\varepsilon
-2\varepsilon ^2/(N-1)$} & \multicolumn{1}{|c|}{$\varepsilon -4\varepsilon
^2/N$} & \multicolumn{1}{c|}{$\varepsilon -2\varepsilon ^2$} & $8/3\pi ^2N$
& $1-32/\pi ^2N$ \\ \cline{5-6}\cline{1-4}
$\nu $ & \multicolumn{1}{c|}{$1/\varepsilon -1/(N-1)$} & 
\multicolumn{1}{|c|}{$1/\varepsilon -2/N$} & $1/\varepsilon -1$ & $1-32/3\pi
^2N$ & $1-48/\pi ^2N$ \\ \hline\hline
\end{tabular}
$

The value of the exponent $\nu $ for the $CP^{N-1}$ model was obtained
earlier from the one-particle excitation spectrum \cite{Ma,Hikami1,Hikami2}.
However, only this exponent can be calculated in such a way. To find a
second exponent and thereby all the other ones, an above-performed explicit
calculation of two-particle Green functions is necessary.

One can see that for $d=2+\varepsilon ,$ $N=2$ our results in the $%
CP^{N-1}\, $model coincide with the corresponding RG values. The results in
the $O(3)$ model differ from those in the $CP^1$ model. This is due to that
the true parameter of $1/N$-expansion in the $O(N)$ model is in fact $%
1/(N-2),$ so that higher order terms are needed to transform powers of $1/N$
into $1/(N-2).$ A similar situation takes place for $d=2$ where higher-order
terms are needed to correct preexponential factor in the correlation length
within the $1/N$ expansion for the $O(N)$ model \cite{Chubukov}, but not for
the $CP^{N-1}$ model \cite{Starykh}.

At $d=3$ the leading $1/N-$corrections to critical exponents are too large
for $N=2$ , so that we obtain negative values of critical exponents.
Apparently, high-order terms are required to eliminate this difficulty. Thus
the mean field approximation \cite{Arovas,ArovasBook}, which describes
satisfactorily low-dimensional systems, is not too convenient starting point
for $3d$ magnetic systems. At the same time, in the $O(N)$-model the $1/N$
corrections are small for $N=3$ and yield physically reasonable results (see 
\cite{MaBook}).

At $d=4-\varepsilon ,$ the propagators $\Pi ^{-1},$ $\Pi _{\mu \nu }^{-1}$
contain the small factor $1/A_d\sim \varepsilon .$ Therefore the corrections
to the critical exponents (\ref{Beta}), (\ref{Eta}) are small. At the same
time, the values of mean-field critical exponents ($\beta =1,$ $\eta =2$)
are in drastic discrepancy with their values $\beta =1/2-O(\varepsilon ),$ $%
\eta =O(\varepsilon ^2)$ calculated by renormalization group approach in the 
$\phi ^4$ model \cite{MaBook}. Taking into account several terms in $1/N-$%
expansion cannot change the situation in this case. Thus summation of an
infinite sequence of diagrams is necessary to correct critical exponents at $%
d=4.$ The situation for $d=4-\varepsilon $ is also complicated and not quite
clear. It was shown by using the $\varepsilon -$ expansion \cite{Ma,Hikami1}
that to first order in $\varepsilon $ the second-order phase transition in
the $CP^{N-1}$ model exists in a standard sense only at $2N>365,$ and at
smaller $N$ the critical exponents turn out to be non-real. This contradicts
to the fact of existence of a second-order transition in the Heisenberg
model ($N=2$).

To conclude, we have calculated critical exponents in the $CP^{N-1}$ model
to first order in $1/N$ starting from the mean-field approximation \cite
{Arovas}. For $d=2+\varepsilon $ the results agree with the scaling analysis 
\cite{Brezin} and the corrections are small. This confirms the fact that the
mean field approximation works well for low-dimensional magnets. For $%
d=4-\varepsilon ,$ our approach, which is based on the $1/N$ expansion,
turns out to be poor. In the intermediate case $d=3$ one has to investigate
next-order corrections to obtain physically satisfactory results.

The work was supported in part by Grant 96-02-1600 from the Russian Basic
Research Foundation.

\newpage 

Fig.1. $1/N$ corrections to the one-particle Green's function of the field $%
z $. Solid line is the bare Green's function, dashed line is the propagator $%
\Pi ^{-1}$ of the field $\lambda ,$ and wavy line is the propagator $\Pi
_{\mu v}^{-1}$ of the field$A.$

Fig.2. $1/N$ corrections to the longitudinal non-uniform susceptibility.


\begin{references}
\bibitem{ArovasBook}  A.Arovas, {\it Interacting\thinspace \thinspace
Electrons\thinspace \thinspace and\thinspace \thinspace Quantum\thinspace
\thinspace Magnetism}, Springer-Verlag,New-York, 1994.

\bibitem{SphModel}  R.J.Baxter, {\it Exactly Solved Models in Statistical
Mechanics,} Academic Press, 1982; T.Vojta, Phys.Rev.B53, 711 (1995)

\bibitem{Hikami1}  S.Hikami, Progr.Theor.Phys. 62, 226 (1979).

\bibitem{Hikami2}  S.Hikami, Progr.Theor.Phys. 64, 1425 (1980).

\bibitem{Arovas}  P.Arovas and A.Auerbach, Phys.Rev.B 38, 316 (1988).

\bibitem{Yoshioka}  D. Yoshioka, J.Phys.Soc.Jpn 58, 3933 (1989).

\bibitem{Takahashi}  M.Takahashi, Phys.Rev.B 40, 2494 (1989).

\bibitem{Our1}  V.Yu.Irkhin, A.A.Katanin, and M.I.Katsnelson, Phys.Lett.A
157, 295 (1991); Fiz.Metallov Metalloved.79, N1, 65 (1995).

\bibitem{Our2}  V.Yu.Irkhin, A.A.Katanin, and M.I.Katsnelson,
J.Phys.:Cond.Mat. 4, 5227 (1992).

\bibitem{Halperin}  S. Chakravarty, B.I.Halperin, and D.R. Nelson
Phys.Rev.B. 39 2344 (1989).

\bibitem{Starykh}  O.A.Starykh, Phys.Rev B 50, 16428 (1994).

\bibitem{Read}  N.Read and S.Sachdev, Phys.Rev.B42 4568 (1990).

\bibitem{CP}  M.Luscher, Phys.Letters, 78B, 465 (1978); A. D'Adda, P.Di
Vecchia, and M. Luscher, Nucl. Phys. B 146, 63 (1978).

\bibitem{MaBook}  S.-K. Ma,{\it \ Modern Theory of Critical Phenomena}
(Benjamin-Cummings, Reading, 1976).

\bibitem{Brezin}  E.Brezin and J.Zinn-Justin, Phys.Rev.B 14, 3110 (1976)

\bibitem{Ma}  B.I.Halperin, T.C.Lubensky, and S.-K. Ma, Phys.Rev.Lett 32,
292 (1974).

\bibitem{Chubukov}  A.V.Chubukov, S.Sachdev, and J.Ye, Phys.Rev.B 49, 11919
(1994).
\end{references}
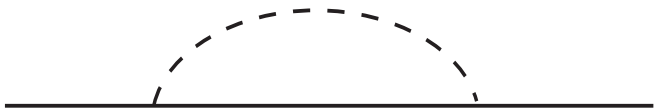
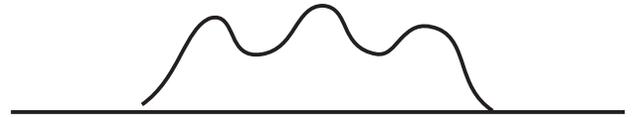
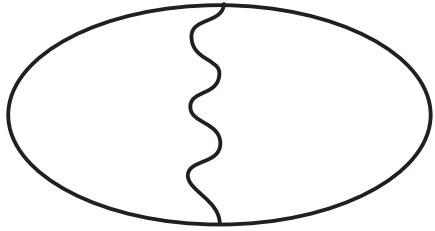
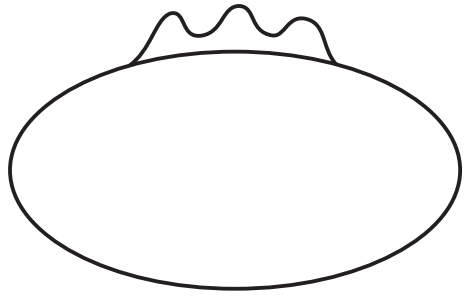
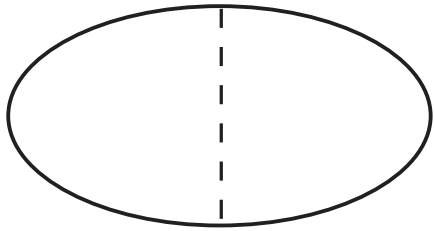
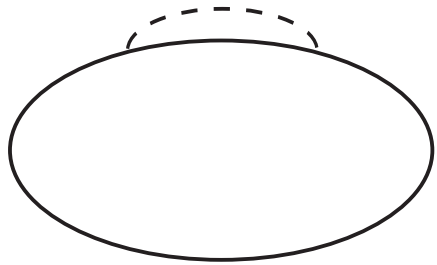
\end{document}